\documentclass{article}

\usepackage[utf8]{inputenc}
\usepackage[T1]{fontenc}
\usepackage{arxiv}
\usepackage{amsmath,amsfonts,amssymb}
\usepackage{graphicx}
\usepackage{float}
\usepackage{booktabs}
\usepackage[numbers]{natbib}
\usepackage[colorlinks=true, allcolors=blue]{hyperref}
\usepackage{authblk}
\usepackage{siunitx}
\usepackage{tikz}
\usetikzlibrary{positioning, arrows.meta}

\title{Visualising Parker Weighting in Short-Scan Cone-Beam Micro-CT: A Practical Reference}
\renewcommand{\shorttitle}{Visualising Parker Weighting in Short-Scan Micro-CBCT}

\author[a]{Falk L. Wiegmann}
\author[a,b]{Nancy L. Ford\thanks{Corresponding author: nlford@dentistry.ubc.ca}}
\affil[a]{Department of Physics and Astronomy, The University of British Columbia, Vancouver BC, Canada}
\affil[b]{Department of Oral Biological and Medical Sciences, The University of British Columbia, Vancouver BC, Canada}

\begin{document}

\maketitle

\begin{abstract}
Short-scan FDK reconstruction is widely used in preclinical cone-beam micro-CT because it reduces scan time and radiation dose, and because the large volume sizes typical of micro-CT make iterative methods impractical for routine use. Short scans, however, introduce non-uniform data redundancy that must be corrected by Parker weighting to avoid directional shading artefacts. This note provides a visual and quantitative summary of Parker weighting as implemented for the eXplore CT~120 scanner. We illustrate the weight maps in the detector and sinogram domains, demonstrate the shading artefacts that arise without correction on both an image quality phantom and an \textit{in vivo} mouse lung, and show via MTF, NPS, and detectability analysis that Parker weighting corrects HU inaccuracies without degrading image quality. No new methods are introduced; the aim is to serve as a concise practical reference for groups implementing or evaluating short-scan FDK pipelines.
\end{abstract}

\section{Introduction}\label{sec:introduction}

In preclinical micro-CT, image volumes are large and iterative reconstruction remains prohibitively slow for routine use, so the Feldkamp--Davis--Kress (FDK) algorithm continues to be the workhorse for cone-beam reconstruction. Short-scan FDK, which acquires projections over an arc of just $\pi + 2\gamma_m$ rather than a full rotation, is especially prevalent because it reduces scan time and radiation dose---both critical concerns in longitudinal small-animal studies. Short-scan acquisition, however, introduces non-uniform data redundancy near the arc boundaries, and without correction this produces directional shading artefacts in the reconstructed image. Parker~\cite{parker_weights_1982} proposed smooth weighting functions that compensate for this redundancy, and the approach is well established in the literature.

This note does not introduce new theory. Rather, it provides a concise visualisation and summary of Parker weighting as applied to cone-beam micro-CT: what the weights look like in the detector and sinogram domains, what happens to the reconstruction when they are omitted, and whether their application affects image quality metrics. Our aim is to make the practical effect of Parker weighting explicit, so that it can serve as a reference for groups implementing or evaluating short-scan FDK pipelines.

\newpage
\section{Background}\label{sec:background}

\subsection{Short-Scan Geometry and Data Redundancy}\label{sec:short_scan}

In circular cone-beam CT (CBCT), the X-ray source traverses a circular orbit around the object. A full $360^{\circ}$ rotation samples every ray path twice: for any ray at gantry angle $\beta$ and fan angle $\gamma$, a conjugate ray measuring the same line integral exists at angle $\beta' = \beta + \pi + 2\gamma$~\cite{parker_weights_1982}. While this redundancy can be exploited for noise reduction, it is not necessary for a complete reconstruction. The minimum angular range required for exact reconstruction in fan-beam geometry is $\pi + 2\gamma_m$, where $\gamma_m$ is the half-fan angle of the detector---commonly referred to as a \emph{short scan}~\cite{parker_weights_1982, kak_slaney_1988}.

Short-scan acquisition is standard in many preclinical micro-CT systems because it reduces scan time, lowers radiation dose, and minimises motion artefacts. However, it introduces a complication: within the short-scan arc, some ray paths are sampled twice (at conjugate angles) while others are sampled only once. Specifically, a ray at $(\beta, \gamma)$ has a conjugate partner within the scan range only if $\beta + 2\gamma \leq 2\gamma_m$---a condition met exclusively for projections near the start and end of the arc, where a ray from one boundary has its conjugate near the opposite boundary. Projections in the middle of the arc have no conjugate partners. If all projections are weighted equally during backprojection, the doubly-sampled rays contribute twice, producing intensity inhomogeneities---shading artefacts that appear as brightness variations across the reconstructed image.

\begin{figure}[H]
    \begin{center}
    \begin{tikzpicture}[scale=1.1]

    \def\R{3.5}          
    \def\objR{0.9}       
    \def\gammaM{6.5}     
    \def\gammaRay{5}     
    \def\betaA{2}        

    \pgfmathsetmacro{\arcSpan}{180 + 2*\gammaM}           
    \pgfmathsetmacro{\betaB}{\betaA + 180 + 2*\gammaRay}  

    \draw[gray, thick] (90:{\R}) arc[start angle=90, end angle={90 - \arcSpan}, radius=\R];
    \draw[gray, thin] (90:{\R-0.15}) -- (90:{\R+0.15});
    \draw[gray, thin] ({90-\arcSpan}:{\R-0.15}) -- ({90-\arcSpan}:{\R+0.15});
    \node[gray, font=\scriptsize, anchor=south] at (90:{\R+0.2}) {$0^{\circ}$};
    \node[gray, font=\scriptsize, anchor=north east] at ({90-\arcSpan}:{\R+0.2}) {$193^{\circ}$};
    \draw[gray, dashed, thin] ({90-\arcSpan}:{\R}) arc[start angle={90-\arcSpan}, end angle={90-360}, radius=\R];

    \fill[pink!30] (0,0) circle (\objR);
    \draw[pink!70!red, thick] (0,0) circle (\objR);
    \node[font=\scriptsize, text=pink!70!red, anchor=west] at ({\objR + 0.08}, 0) {Object};
    \fill[black] (0,0) circle (1.5pt);

    \pgfmathsetmacro{\sAx}{\R * cos(90 - \betaA)}
    \pgfmathsetmacro{\sAy}{\R * sin(90 - \betaA)}

    \fill[red!80!black] (\sAx, \sAy) circle (3pt);
    \node[font=\small\bfseries, text=red!80!black, anchor=south west]
          at (\sAx + 0.08, \sAy + 0.08) {$S_1$};

    \pgfmathsetmacro{\cAdx}{-\sAx / \R}
    \pgfmathsetmacro{\cAdy}{-\sAy / \R}
    \pgfmathsetmacro{\cAex}{\sAx + 4.0 * \cAdx}
    \pgfmathsetmacro{\cAey}{\sAy + 4.0 * \cAdy}
    \draw[red!80!black, thin, dashed, opacity=0.4] (\sAx, \sAy) -- (\cAex, \cAey);

    \pgfmathsetmacro{\rayAdx}{\cAdx * cos(-\gammaRay) - \cAdy * sin(-\gammaRay)}
    \pgfmathsetmacro{\rayAdy}{\cAdx * sin(-\gammaRay) + \cAdy * cos(-\gammaRay)}

    \pgfmathsetmacro{\sBx}{\R * cos(90 - \betaB)}
    \pgfmathsetmacro{\sBy}{\R * sin(90 - \betaB)}

    \pgfmathsetmacro{\midx}{(\sAx + \sBx) / 2}
    \pgfmathsetmacro{\midy}{(\sAy + \sBy) / 2}

    \draw[red!80!black, thick, -{Stealth[length=4pt]}]
          (\sAx, \sAy) -- (\midx, \midy);

    \pgfmathsetmacro{\arcRA}{2.2}
    \pgfmathsetmacro{\angCA}{atan2(\cAdy, \cAdx)}
    \pgfmathsetmacro{\angRA}{atan2(\rayAdy, \rayAdx)}
    \draw[red!80!black, thin, <->]
          ([shift=(\angRA:\arcRA)] \sAx, \sAy)
          arc[start angle=\angRA, end angle=\angCA, radius=\arcRA];
    \pgfmathsetmacro{\angMidA}{(\angCA + \angRA) / 2}
    \node[font=\scriptsize, text=red!80!black, anchor=west]
          at ([shift=(\angMidA:{\arcRA + 0.08})] \sAx, \sAy) {$\gamma$};

    \draw[gray, thin, dashed] (0, 0) -- (\sAx, \sAy);
    \node[gray, font=\scriptsize, anchor=south] at (0.35, {\R + 0.4})
          {$\beta = 2^{\circ}$};
    \draw[gray, thin, ->] (0.35, {\R + 0.3}) -- (0.15, {\R + 0.08});

    \fill[green!50!black] (\sBx, \sBy) circle (3pt);
    \node[font=\small\bfseries, text=green!50!black, anchor=north west]
          at (\sBx + 0.12, \sBy - 0.12) {$S_2$};

    \pgfmathsetmacro{\cBdx}{-\sBx / \R}
    \pgfmathsetmacro{\cBdy}{-\sBy / \R}
    \pgfmathsetmacro{\cBex}{\sBx + 4.0 * \cBdx}
    \pgfmathsetmacro{\cBey}{\sBy + 4.0 * \cBdy}
    \draw[green!50!black, thin, dashed, opacity=0.4] (\sBx, \sBy) -- (\cBex, \cBey);

    \pgfmathsetmacro{\rayBdx}{\cBdx * cos(\gammaRay) - \cBdy * sin(\gammaRay)}
    \pgfmathsetmacro{\rayBdy}{\cBdx * sin(\gammaRay) + \cBdy * cos(\gammaRay)}

    \draw[green!50!black, thick, -{Stealth[length=4pt]}]
          (\sBx, \sBy) -- (\midx, \midy);

    \pgfmathsetmacro{\arcRB}{2.2}
    \pgfmathsetmacro{\angCB}{atan2(\cBdy, \cBdx)}
    \pgfmathsetmacro{\angRB}{atan2(\rayBdy, \rayBdx)}
    \draw[green!50!black, thin, <->]
          ([shift=(\angCB:\arcRB)] \sBx, \sBy)
          arc[start angle=\angCB, end angle=\angRB, radius=\arcRB];
    \pgfmathsetmacro{\angMidB}{(\angCB + \angRB) / 2}
    \node[font=\scriptsize, text=green!50!black, anchor=west]
          at ([shift=(\angMidB:{\arcRB + 0.08})] \sBx, \sBy) {$-\gamma$};

    \draw[gray, thin, dashed] (0, 0) -- (\sBx, \sBy);
    \node[gray, font=\scriptsize, anchor=north east] at ({\sBx - 0.2}, {\sBy + 0.35})
          {$\beta' = 192^{\circ}$};

    \node[font=\scriptsize, align=left, anchor=north west] at (3.0, -2.6) {%
        \textcolor{red!80!black}{\rule{8pt}{2pt}} Ray from $S_1$ at $(\beta,\, \gamma)$\\[2pt]
        \textcolor{green!50!black}{\rule{8pt}{2pt}} Conjugate ray from $S_2$ at $(\beta',\, {-}\gamma)$\\[2pt]
        $\beta' = \beta + \pi + 2\gamma$\\[2pt]
        Both rays traverse the same line integral.};

    \end{tikzpicture}
    \end{center}
    \caption{Conjugate ray geometry for the eXplore CT~120 short-scan acquisition ($193^{\circ}$, $\gamma_m = 6.5^{\circ}$). Source $S_1$ at gantry angle $\beta = 2^{\circ}$ emits a ray at fan angle $\gamma = 5^{\circ}$; source $S_2$ at the conjugate angle $\beta' = \beta + \pi + 2\gamma = 192^{\circ}$ emits a ray at fan angle $-\gamma$ that traverses the same path through the object. Such conjugate pairs exist only for rays near the arc boundaries where $\beta + 2\gamma \leq 2\gamma_m$, creating the non-uniform data redundancy that Parker weighting corrects. Solid grey arc: scanned range; dashed: unscanned.}
    \label{fig:conjugate_rays}
\end{figure}
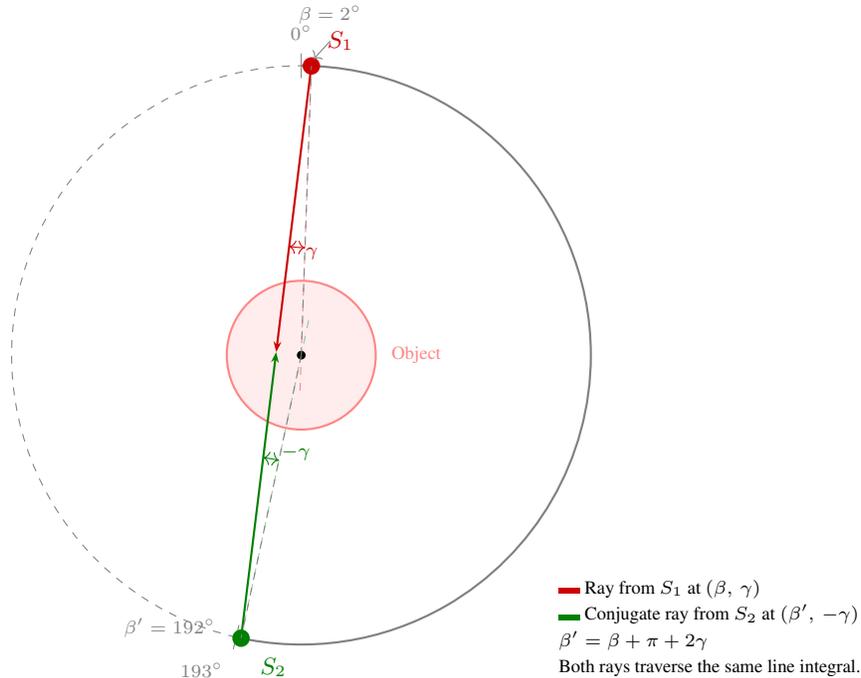

Figure~\ref{fig:conjugate_rays} illustrates this geometry. A ray from source position $S_1$ at gantry angle $\beta$ with fan angle $\gamma$ traverses the same path through the object as a ray from source position $S_2$ at gantry angle $\beta' = \beta + \pi + 2\gamma$ with fan angle $-\gamma$. These conjugate rays measure the same line integral; without correction, both contribute fully to the reconstruction, producing the shading artefacts described above.

\newpage

\subsection{Parker Weighting}\label{sec:parker_weights}

Parker~\cite{parker_weights_1982} proposed a smooth weighting function that assigns fractional weights to each ray such that conjugate ray pairs always sum to a constant. For a short scan spanning gantry angles $\beta \in [0, \, \pi + 2\gamma_m]$, the Parker weight $w(\beta, \gamma)$ is defined piecewise as:
\begin{equation}\label{eq:parker}
w(\beta, \gamma) =
\begin{cases}
\sin^2\!\left(\dfrac{\pi\,\beta}{4(\gamma_m - \gamma)}\right) & \text{if } 0 \leq \beta < 2(\gamma_m - \gamma), \\[10pt]
1 & \text{if } 2(\gamma_m - \gamma) \leq \beta \leq \pi - 2\gamma, \\[10pt]
\sin^2\!\left(\dfrac{\pi\,(\pi + 2\gamma_m - \beta)}{4(\gamma_m + \gamma)}\right) & \text{if } \pi - 2\gamma < \beta \leq \pi + 2\gamma_m, \\[10pt]
0 & \text{otherwise},
\end{cases}
\end{equation}
where $\beta$ is the gantry angle measured from the start of the scan and $\gamma$ is the signed fan angle of a given detector element ($\gamma = 0$ at the central ray, $|\gamma| \leq \gamma_m$). The $\sin^2$ transitions ensure that the weights are smooth at the boundaries: $w = 0$ at the start and end of the scan for peripheral rays, rising to $w = 1$ over the central portion of the arc where no redundancy exists. Conjugate rays satisfy $w(\beta, \gamma) + w(\beta + \pi + 2\gamma, -\gamma) = 1$, ensuring uniform weighting across all ray paths.\footnote{Some formulations define the weights to sum to 2 rather than 1, absorbing a factor of $\frac{1}{2}$ into the backprojection normalisation. We use the sum-to-one convention throughout.}

\begin{figure}[H]
    \begin{center}
    \includegraphics[width=\textwidth]{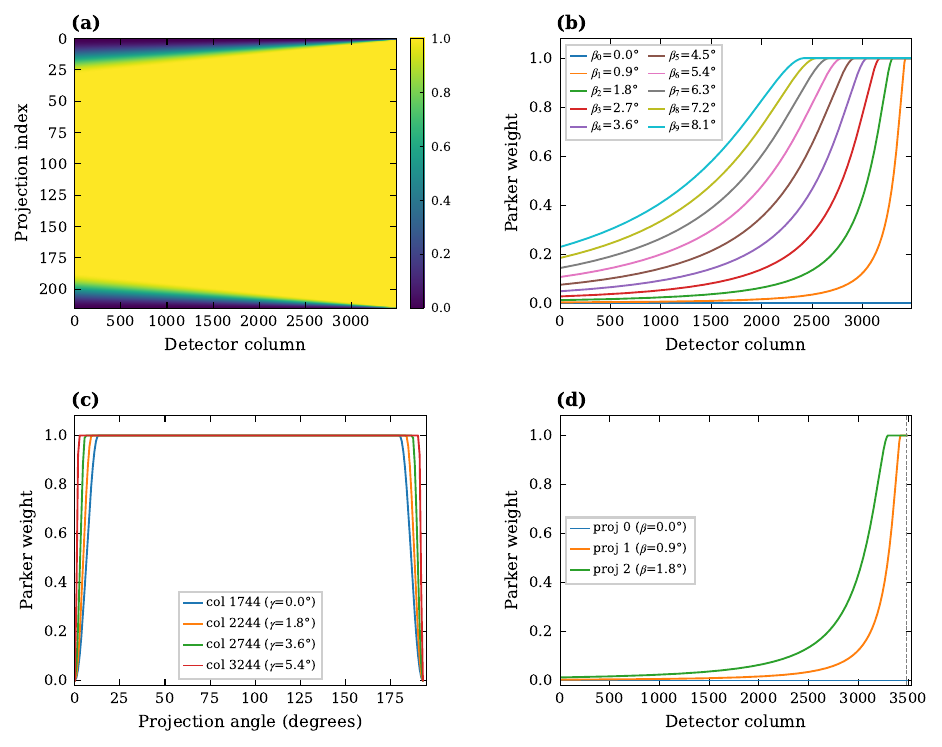}
    \end{center}
    \caption{Parker weighting for the eXplore CT~120 short-scan geometry ($193^{\circ}$). (a)~Two-dimensional weight map across detector columns and projection indices. (b)~Weight profiles across detector columns for the first ten projection angles $\beta_0$--$\beta_9$. (c)~Weight profiles across projection angles for selected detector columns at different fan angles $\gamma$. (d)~Weight profiles for the first three projections across the full detector width.}
    \label{fig:parker_weights}
\end{figure}

Figure~\ref{fig:parker_weights} illustrates the Parker weight map computed for the eXplore CT~120 short-scan geometry ($193^{\circ}$). Panel~(a) shows the two-dimensional weight map across all detector columns and projection indices: the bulk of the map is uniformly 1.0, with the $\sin^2$ ramp-up and ramp-down regions visible at early and late projection angles for peripheral detector columns. Panels~(b--d) show cross-sections of this weight map. Panel~(b) plots the weight as a function of detector column for the first ten projection angles, illustrating how the ramp onset shifts with gantry angle. Panel~(c) shows the weight as a function of projection angle for selected detector columns at different fan angles $\gamma$: peripheral columns ($\gamma = 5.4^{\circ}$) reach $w = 1$ within just a few projections because their ramp-up region $2(\gamma_m - \gamma)$ is narrow, while central columns ($\gamma \approx 0^{\circ}$) ramp more gradually over a wider angular range. Panel~(d) shows the weight across the full detector width for the first three projections, highlighting the asymmetric structure at the scan-arc start.

Parker's original formulation was derived for two-dimensional fan-beam CT. Wesarg et al.~\cite{wesarg_parker_2002} generalised it by deriving a class of smooth weighting functions valid for arbitrary scan angles from $\pi + 2\gamma_m$ to $2\pi$, with the original Parker weights as a special case, and noted that the extension to three-dimensional cone-beam geometry via the FDK algorithm~\cite{feldkamp_fdk_1984} is straightforward. In practice, the fan-beam weights are applied identically to each detector row based on the horizontal fan angle, with the cone angle handled by the FDK approximation. Silver~\cite{silver_redundant_2000} separately addressed the intermediate case where the scan arc falls between $\pi + 2\gamma_m$ and $2\pi$, introducing a virtual fan angle to manage partial redundancy.

\subsection{The FDK Pipeline}\label{sec:fdk_pipeline}

The Feldkamp--Davis--Kress (FDK) algorithm~\cite{feldkamp_fdk_1984} reconstructs a three-dimensional volume from cone-beam projections acquired along a circular trajectory. For a flat-panel detector, each projection $p(\beta, u, v)$ at gantry angle $\beta$ with detector coordinates $(u, v)$ is processed through the following steps:
\begin{enumerate}
    \item \textbf{Pre-processing}: dark-current subtraction, flood-field normalisation, and logarithmic transform to convert raw detector intensities to line integrals of attenuation.
    \item \textbf{Cosine weighting}: multiplication by $D / \sqrt{D^2 + u^2 + v^2}$, where $D$ is the source-to-detector distance, to account for the varying path lengths in cone-beam geometry.
    \item \textbf{Parker weighting}: for short-scan acquisitions, multiplication of each projection--detector element by $w(\beta, \gamma)$ as defined in Equation~\ref{eq:parker}.
    \item \textbf{Ramp filtering}: one-dimensional convolution of each detector row with a ramp-type filter kernel $h(u)$ whose frequency response is $H(f) = |f|$, optionally apodised by a window function.
    \item \textbf{Backprojection}: three-dimensional voxel-driven backprojection with distance weighting.
    \item \textbf{HU calibration}: conversion from linear attenuation to Hounsfield units using air and water reference values.
\end{enumerate}

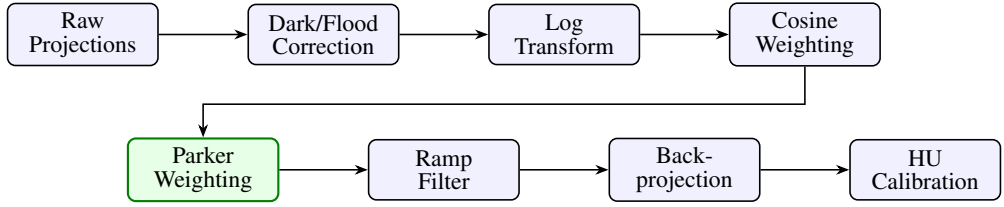
\begin{figure}[H]
    \begin{center}
    \begin{tikzpicture}[
        box/.style={
            draw, rounded corners=3pt, minimum height=0.8cm,
            minimum width=2.0cm, align=center, font=\small,
            fill=blue!6, line width=0.6pt
        },
        parkerbox/.style={
            draw, rounded corners=3pt, minimum height=0.8cm,
            minimum width=2.0cm, align=center, font=\small,
            fill=green!10, line width=0.8pt, draw=green!50!black
        },
        arrow/.style={-{Stealth[length=5pt]}, line width=0.6pt},
    ]

    \node[box] at (0, 0) (raw) {Raw\\[-1pt]Projections};
    \node[box] at (3.2, 0) (preproc) {Dark/Flood\\[-1pt]Correction};
    \node[box] at (6.4, 0) (log) {Log\\[-1pt]Transform};
    \node[box] at (9.6, 0) (cosine) {Cosine\\[-1pt]Weighting};

    \draw[arrow] (raw) -- (preproc);
    \draw[arrow] (preproc) -- (log);
    \draw[arrow] (log) -- (cosine);

    \node[parkerbox] at (1.6, -1.8) (parker) {Parker\\[-1pt]Weighting};
    \node[box] at (4.8, -1.8) (ramp) {Ramp\\[-1pt]Filter};
    \node[box] at (8.0, -1.8) (bp) {Back-\\[-1pt]projection};
    \node[box] at (11.2, -1.8) (hu) {HU\\[-1pt]Calibration};

    \draw[arrow] (cosine.south) -- ++(0, -0.5) -| (parker.north);
    \draw[arrow] (parker) -- (ramp);
    \draw[arrow] (ramp) -- (bp);
    \draw[arrow] (bp) -- (hu);

    \end{tikzpicture}
    \end{center}
    \caption{FDK reconstruction pipeline for short-scan cone-beam micro-CT. Parker weighting (highlighted) is applied after cosine weighting and before ramp filtering.}
    \label{fig:pipeline}
\end{figure}

\noindent Figure~\ref{fig:pipeline} summarises this pipeline.

\section{Methods}\label{sec:methods}

All data were acquired on the eXplore CT~120 micro-CT scanner (Trifoil Imaging, USA) equipped with a flat-panel detector of $3500 \times 2300$ pixels. The scanner performs a short-scan acquisition over $193^{\circ}$ ($180^{\circ} + 2\gamma_m$) with 220 projections at \SI{80}{kVp} and \SI{40}{mA}. Two specimens were used to evaluate the effect of Parker weighting: (1)~an mCTP~610 image quality phantom (Shelley Medical Imaging Technologies, Canada), providing a geometrically simple object with known uniform regions for quantitative analysis; and (2)~an \textit{in vivo} mouse lung, representing a typical preclinical imaging target with complex anatomical structures.

Our FDK reconstruction pipeline was implemented from scratch following the steps described in Section~\ref{sec:fdk_pipeline}. The flat-panel detector has a pixel pitch of \SI{28.4}{\micro m}. Volumes were reconstructed on a $1247 \times 1247 \times 933$ grid with isotropic \SI{75}{\micro m} voxels.

To demonstrate the effect of Parker weighting, each scan was reconstructed twice with all pipeline parameters held constant:
\begin{itemize}
    \item[(a)] Without Parker weighting (all projections weighted equally).
    \item[(b)] With Parker weighting applied according to Equation~\ref{eq:parker}.
\end{itemize}
A voxel-wise residual map (a)~$-$~(b) was computed for each specimen to visualise and quantify the spatial distribution of intensity corrections introduced by the weighting.

To assess whether Parker weighting affects image quality beyond the removal of shading artefacts, three physics-based metrics were computed from the image quality phantom reconstructions. The modulation transfer function (MTF) was measured using the slanted-edge method applied to the air--water interface of a rectangular insert (edge angle $\approx 3.9^{\circ}$) and reports the spatial frequency at which the response falls to 10\% of its peak (MTF$_{10}$). The noise power spectrum (NPS) was computed from eight square regions of interest (ROIs) arranged in a ring within a homogeneous water region of the phantom and averaged to characterise the noise texture and magnitude as a function of spatial frequency. Finally, the non-prewhitening (NPW) matched-filter detectability index $d'$ was derived from the MTF and NPS to provide a task-based measure of low-contrast detectability for disc-shaped signals of varying diameter at contrast levels of $\Delta C = 100$ and \SI{500}{HU}.

\newpage
\section{Results}\label{sec:results}

\subsection{Parker Weighting in the Sinogram Domain}\label{sec:sinogram_domain}

\begin{figure}[H]
    \begin{center}
    \includegraphics[width=\textwidth]{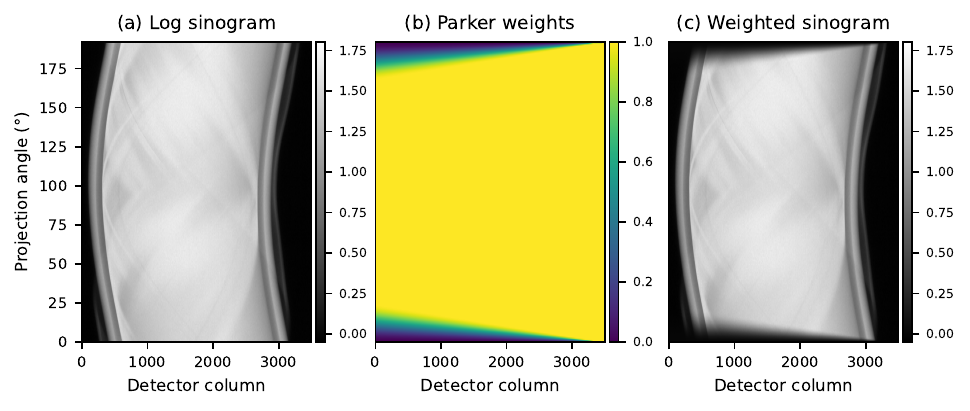}
    \end{center}
    \caption{Parker weighting applied in the sinogram domain for a representative detector row. (a)~Log-transformed sinogram showing line integrals of attenuation. (b)~Parker weight map $w(\beta, \gamma)$ for the eXplore CT~120 short-scan geometry ($193^{\circ}$). (c)~Weighted sinogram: the product of (a) and (b), with data at the scan-arc boundaries smoothly attenuated.}
    \label{fig:sinogram_stages}
\end{figure}

Figure~\ref{fig:sinogram_stages} shows the effect of Parker weighting on a representative detector row in the sinogram domain. Panel~(a) displays the log-transformed sinogram (line integrals of attenuation) as a function of detector column and projection angle. Panel~(b) shows the corresponding Parker weight map, where the $\sin^2$ transitions at early and late projection angles are clearly visible---particularly at the left edge of the detector (large negative fan angles), where the ramp regions extend over a wider angular range. Panel~(c) shows the weighted sinogram: the product of (a) and (b). The weighting smoothly attenuates data at the scan-arc boundaries, ensuring that conjugate rays contribute proportionally rather than doubly.

\subsection{Effect on Reconstruction Quality}\label{sec:recon_effect}

\begin{figure}[H]
    \begin{center}
    \includegraphics[width=0.9\textwidth]{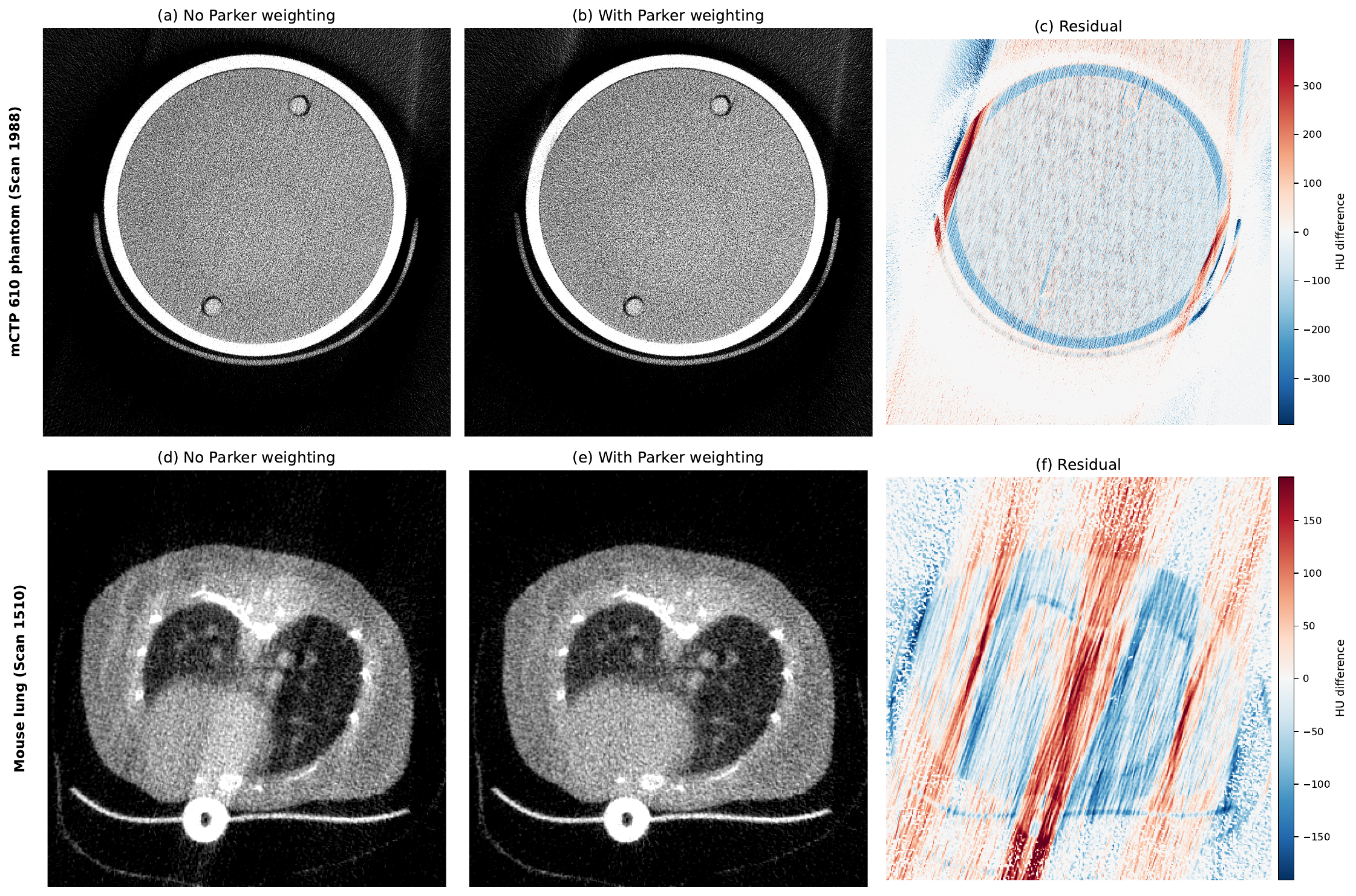}
    \end{center}
    \caption{Effect of Parker weighting on reconstruction quality. Top row: image quality phantom. Bottom row: mouse lung. (a,\,d)~Without Parker weighting. (b,\,e)~With Parker weighting. (c,\,f)~Residual maps (unweighted~$-$~weighted) in HU.}
    \label{fig:parker_comparison}
\end{figure}

Figure~\ref{fig:parker_comparison} presents central-slice reconstructions of both specimens with and without Parker weighting. The top row shows the image quality phantom: panel~(a) without Parker weighting exhibits directional shading artefacts as intensity inhomogeneities, particularly in the lower portion of the phantom and near the outer edges where conjugate ray contributions are strongest; panel~(b) with Parker weighting eliminates these artefacts, producing a more uniform image. The residual map in panel~(c) shows the correction concentrated at the phantom boundary and in the lower region of the image where redundant rays contribute most unevenly. The interior shows a low-amplitude, spatially varying offset, indicating that the shading artefact affects bulk HU values of homogeneous regions, not just edges.

The bottom row shows the mouse lung. The unweighted reconstruction~(d) exhibits a directional intensity gradient and shading across the image. Parker weighting~(e) removes this gradient. The residual map~(f) shows a distinctive directional striping pattern that differs from the ring-like structure seen in the phantom residual, indicating that the spatial distribution of the shading artefact depends on the object's attenuation profile.

\subsection{Quantitative Image Quality Metrics}\label{sec:metrics}

\begin{figure}[H]
    \begin{center}
    \includegraphics[width=\textwidth]{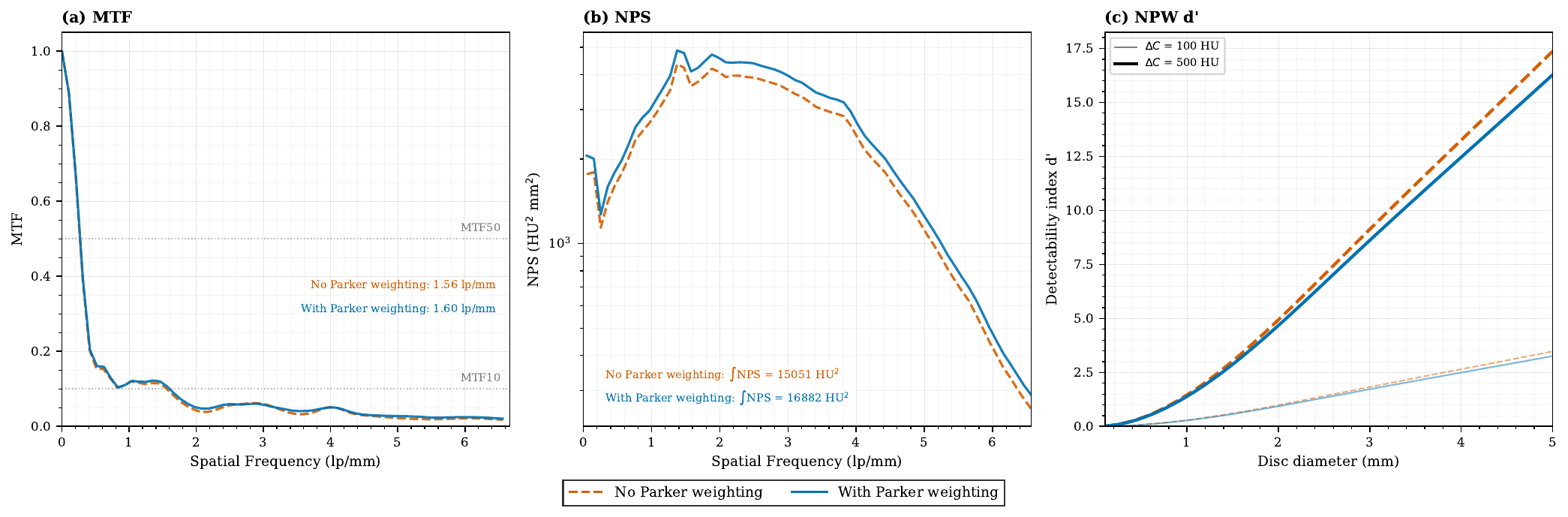}
    \end{center}
    \caption{Image quality metrics for image quality phantom reconstructions with and without Parker weighting. (a)~Modulation transfer function: spatial resolution is essentially unchanged (MTF$_{10}$: 1.60 vs 1.56\,lp/mm). (b)~Noise power spectrum: integrated noise power is generally lower without Parker weighting. (c)~Non-prewhitening detectability index $d'$ for disc signals at $\Delta C = 100$ and \SI{500}{HU}: detectability is similar for both conditions.}
    \label{fig:parker_metrics}
\end{figure}

Figure~\ref{fig:parker_metrics} compares the MTF, NPS, and detectability index for image quality phantom reconstructions with and without Parker weighting. The MTF curves~(a) are nearly identical (MTF$_{10}$: 1.60 vs 1.56\,lp/mm), confirming that spatial resolution is unaffected. The NPS~(b) shows generally lower integrated noise power without Parker weighting, consistent with the shading artefact being a smooth deterministic bias that does not register as noise; its removal reveals the underlying stochastic noise more faithfully. The NPW detectability index $d'$~(c) is similar for both conditions, indicating a negligible effect on task-based detectability.

\section{Discussion}\label{sec:discussion}

The residual maps demonstrate that Parker weighting is not merely a cosmetic correction: omitting it produces quantitatively incorrect HU values that would compromise downstream analyses such as tissue classification or bone density measurement. The two specimens exhibit qualitatively different residual patterns---ring-like at the phantom boundary versus directional striping in the mouse lung---indicating that the artefact's spatial expression depends on the object's attenuation structure. Image quality metrics confirm that correcting this bias does not degrade spatial resolution or task-based detectability; the lower integrated NPS without Parker weighting reflects the smooth deterministic shading being absent from the noise estimate rather than a true noise reduction.

The extension from fan-beam to cone-beam geometry is handled row-by-row via the FDK approximation~\cite{wesarg_parker_2002}, where the cone angle is not explicitly accounted for in the weighting. For the eXplore CT~120, the cone half-angle is modest and this approximation is expected to be adequate. The Parker weight formulation itself is general, though the specific weight maps and correction magnitudes will differ for scanners with different fan angles or scan arcs.

\section{Conclusion}\label{sec:conclusion}

Parker weighting is essential for quantitatively correct short-scan cone-beam CT reconstruction. On the eXplore CT~120, omitting the weighting introduces directional shading artefacts in both a image quality phantom and an \textit{in vivo} mouse lung that produce measurable HU errors---compromising any analysis dependent on accurate attenuation values. Quantitative image quality metrics confirm that Parker weighting achieves this correction without degrading spatial resolution or task-based detectability. The implementation follows the standard formulation of Parker~\cite{parker_weights_1982} extended to cone-beam geometry via the FDK algorithm, and is part of our open-source reconstruction pipeline for the eXplore CT~120 scanner.

\bibliographystyle{unsrtnat}
\bibliography{parker_weighting_paper}

\section*{Acknowledgments}
This work was supported by the BC Lung Foundation.

\section*{Author Contributions}
Falk L Wiegmann and Nancy L Ford contributed to the research direction and conceptualisation. Falk L Wiegmann developed the reconstruction pipeline, performed the analysis, and wrote the manuscript. Nancy L Ford supervised the research, secured funding, provided critical review, and edited the manuscript.

\section*{Competing Interests}
The authors declare no competing interests.

\section*{Data Availability}
The reconstruction pipeline code is publicly available at \url{https://github.com/UBC-Ford-lab/eXplore_CT_120_fdk_reconstruction_algorithm}.
The scan data are available from the corresponding author on reasonable request.

\section*{Ethics Statement}
No animal experiments were conducted as part of this study. The image quality phantom data were acquired exclusively for evaluation purposes. The \textit{in vivo} mouse lung micro-CT data were collected as part of previously published studies~\cite{ford_respiratory_2025, ford_spie_2023} under protocols approved by the University of British Columbia Animal Care Committee (Protocol No.\ A21-0060, approved August 31, 2021) and performed in accordance with the ARRIVE guidelines~\cite{percie_du_sert_arrive_2020}. The imaging data were re-used here for reconstruction evaluation only; no animals were imaged, handled, or subjected to any procedures as part of this work.

\section*{Use of AI Tools}
Claude (Anthropic) was used to assist with manuscript preparation. All AI-generated content was reviewed, verified, and revised by the authors, who take full responsibility for the final manuscript.

\end{document}